\documentclass[twocolumn,showpacs,preprintnumbers,amsmath,amssymb]{revtex4}

\usepackage{graphicx}
\usepackage{dcolumn}
\usepackage{bm}

\begin{document}
\title{Density oscillations in trapped dipolar condensates}

\author{H.-Y. Lu$^1$, H. Lu$^2$, J.-N. Zhang$^1$, R.-Z. Qiu$^1$, H. Pu$^2$ and S. Yi$^1$}
\affiliation{$^1$Key Laboratory of Frontiers in Theoretical Physics, Institute of Theoretical Physics, Chinese Academy of Sciences, Beijing, 100190, China}
\affiliation{$^2$Dept. of Physics and Astronomy, and Rice Quantum Institute, Rice University, Houston, TX 77251, USA}

\begin{abstract}
We investigated the ground state wave function and free expansion of a trapped dipolar condensate. We find that dipolar interaction may induce both biconcave and dumbbell density profiles in, respectively, the pancake- and cigar-shaped traps. On the parameter plane of the interaction strengths, the density oscillation occurs only when the interaction parameters fall into certain isolated areas. The relation between the positions of these areas and the trap geometry is explored. By studying the free expansion of the condensate with density oscillation, we show that the density oscillation is detectable from the time-of-flight image.
\end{abstract}

\date{\today}

\maketitle

\section{Introduction}
Bose-Einstein condensates with dipole-dipole interaction have drawn significant interests over the past few years~\cite{yi1,goral,santos}. In particular, after the experimental observations of dipolar effects in Cr condensates~\cite{stuh} and in spin-1 Rb~\cite{kurn} condensates, the study of dipolar quantum gases is at the forefront of both theoretical and experimental research. Armed with the technique of Feshbach resonance which tunes the short-range $s$-wave scattering length, experimentalists have investigated the stability and the anisotropic collapse of the Cr condensate~\cite{koch,laha}. In addition, dipolar effects have also been observed in condensates of K~\cite{fatt} and Li~\cite{hulet} atoms which only possess small magnetic dipole moments. Owing to the large electric dipole moment of the polar molecules, recent success in making high phase-space-density gas of polar molecules~\cite{ye1,ye2} provides an even more promising platform for the study the dipolar quantum many-body system.

The long-range and anisotropic characters of the dipolar interaction have profound impact on the properties of quantum gases. For example, when loaded into a optical lattice, the long-range repulsive dipole-dipole interaction breaks the translational symmetry of the lattice such that the density wave phases can be formed~\cite{goral2,yi2}. In the presence of vortices~\cite{yi3} or a small perturbation in the trapping potential~\cite{wilson}, the radial density profile of the condensate was found to develop oscillations. Of particular interest is that, even for a singly trapped dipolar condensate, a biconcave density profile can be found in a pancake-shape trap~\cite{ronen2}. In this case, the maximum density occurs at the periphery of the condensate, resembling the shape of a red blood cell. Wilson {\it et al}. have proposed to detect the biconcave density profile by tuning the contact interaction to induce local collapse~\cite{wilson2}.

In the present paper, we systematically investigate the ground state structure and the free expansion dynamics of a harmonically trapped dipolar condensate. For a given trap geometry, we map out the stability diagrams on the parameter plane of the interaction strengths. In addition to the biconcave dipolar condensates in pancake-shaped traps, a dumbbell density profile can be induced in cigar-shaped trap when the dipolar interaction is properly tuned. For simplicity, both biconcave and dumbbell density profiles are referred to as density oscillations in the condensates. On stability diagram, the density oscillation occurs only when the interaction strengths fall into certain isolated areas, which is called as density oscillation island (DOI). We also find the relation between the position of DOI and the trap geometry by fitting the numerical results. Finally, by studying the free expansion of the condensate with density oscillation, we show that the density oscillation can be identified in the time-of-flight image.

We will also consider dipolar condensate trapped in box potentials. We show that the rigid walls of box potentials in general induce more pronounced density oscillations over a much larger region of the parameter space than harmonic potentials.

The remainder of this paper is organized as follows. In Sec.~\ref{stheo}, we briefly outline our model. In Sec.~\ref{snum}, we introduce the numerical method used to compute the dipolar mean-field. The numerical results on both the ground state wave function of harmonically trapped dipolar condensates and the corresponding free expansion dynamics are presented in Sec.~\ref{sresu}. In Sec.~\ref{box}, we discuss the enhanced density oscillation in a box potential. Finally, we conclude in Sec.~\ref{sconc}.

\section{Formulation}\label{stheo}
We consider a trapped condensate of $N$ bosons with dipole moments $d\, \hat{z}$. Particles interact with each other via contact interaction $V_0$ and dipole-dipole interaction $V_d$. The total interaction potential takes the form
\begin{eqnarray}
V({\mathbf r})=c_0\delta({\mathbf r})+fd^2\frac{1-3\cos^2\theta}{|{\mathbf r}|^3},
\end{eqnarray}
where $c_0=4\pi\hbar^2a_{s}/m$ with $m$ being the mass of the particle and $a_s$ being the $s$-wave scattering length tunable through Feshbach resonance, $\theta$ is the polar angle of ${\mathbf r}$, and $f\in[-\frac{1}{2},1]$ is a parameter continuously tunable via an orienting field which rotates rapidly around $z$-axis~\cite{giov}. The inclusion of the factor $f$ makes the dipolar interaction between polarized dipoles tunable. For positive $f$, the dipolar force is repulsive (attractive) if two dipole moments aligned side by side (head to tail). However, the opposite becomes true for negative $f$. Moreover, the trapping potential is assumed to be axially symmetric $$U_{\rm ho}({\mathbf r})=\frac{1}{2}m\omega_\perp^2(x^2+y^2+\lambda^2z^2),$$
where $\omega_\perp$ is the radial frequency and $\lambda$ is trap aspect ratio such that the axially frequency of the trap is $\omega_z=\lambda\omega_\perp$. 

The condensate wave function satisfies the non-local Gross-Pitaevskii equation, which, in dimensionless form, becomes
\begin{eqnarray}
i\frac{\partial \psi({\mathbf r})}{\partial
t}&=&\left[-\frac{1}{2}\nabla^2+\frac{1}{2}(x^2+y^2+\lambda^2z^2)+g|\psi({\mathbf
r})|^2\right.\nonumber\\
&&\left.+D\int d{\mathbf
r}'\frac{1-3\cos^2\theta}{|{\mathbf r}-{\mathbf
r}'|^3}|\psi({\mathbf r}')|^2\right]\psi({\mathbf
r}),\label{gp3d}
\end{eqnarray}
where $g=4\pi Na/a_\perp$ and $D=fd^2/(\hbar\omega_\perp a_\perp^3)$ are the dimensionless parameters, characterizing the strengths of the contact and dipolar interactions, respectively. The dimensionless units adopted here are $a_\perp=\sqrt{\hbar/(m\omega_\perp)}$ for length, $\hbar\omega_\perp$ for energy, $\omega_\perp^{-1}$ for time, and $\sqrt{N/a_\perp^3}$ for wave function. In Eq.~(\ref{gp3d}), the control parameters of the system are reduced to $\lambda$, $g$, and $D$. We shall show how the condensate stability and the structure of the wave function depend on these parameters.

\section{Numerical method}\label{snum}
\begin{figure}
\centering
\includegraphics[width=3.in]{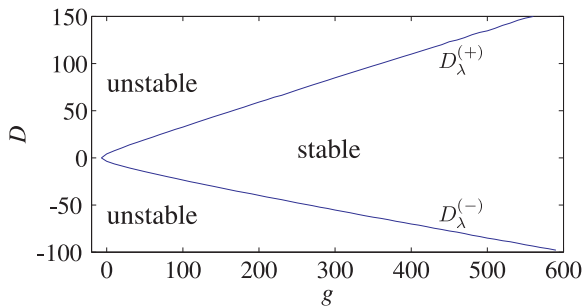}
\caption{The stability diagram for $\lambda=1$.}
\label{cril01}
\end{figure}

Following the standard procedure, the numerical solution of the ground state wave function can be obtained by evolving Eq.~(\ref{gp3d}) in imaginary time. Compared to the Gross-Pitaevskii equation with only contact interaction, the difficulty of solving Eq. (\ref{gp3d}) lies at the calculating the mean-field potential of the dipole-dipole interaction
\begin{eqnarray}
\Phi_d({\mathbf r})&=&\int d{\mathbf r}'V_d({\mathbf r}-{\mathbf r}')n({\mathbf r}')\nonumber\\
&=&{\cal F}^{-1}\left[\widetilde V_d({\mathbf k})\tilde n({\mathbf k})\right],\label{phidd}
\end{eqnarray}
where ${\cal F}^{-1}$ denotes the inverse Fourier transform, $n({\mathbf r})=|\psi({\mathbf r})|^2$ is the density of the condensate with $\tilde n({\mathbf k})$ being its Fourier transform, and $\widetilde V_d({\mathbf k})=\frac{4\pi}{3}fd^2(3\cos^2\theta_k-1)$ is the Fourier transform of the dipolar interaction potential with $\theta_k$ being the polar angle of ${\mathbf k}$. From the second line of Eq. (\ref{phidd}), it becomes apparent that $\Phi_d$ can be efficiently calculated in momentum space using fast Fourier transform (FFT)~\cite{goral}. Moreover, taking advantage of the cylindric symmetry of the system, Ronen {\it et al}.~\cite{ronen} further reduce the 2D Fourier transform in $(x,y)$ plane into an 1D Hankel transform of order $0$ for the ground state wave function by integrating over the azimuthal variable. This procedure greatly simplifies the 3D calculation to a 2D one in the $\rho$-$z$ plane.

The fact that FFT implicitly treats the system as a periodic one makes our system a 3D periodic lattice of condensates. Therefore, the dipolar interactions between the artificial periodic copies of condensates introduce error to the mean-field potential $\Phi_d({\mathbf r})$. To solve this problem, Ronen {\it et al}.~\cite{ronen} suggest to truncate the dipolar interaction such that $V_d$ is nonzero only within a sphere of radius $R$, i.e.,
\begin{eqnarray}
V_d^{\rm (s)}({\mathbf r})=\left\{\begin{array}{cl}
fd^2(1-3\cos^2\theta)/r^3,&r<R,\\
0,&{\rm otherwise}.
\end{array}
\right.\label{cutsph}
\end{eqnarray}

\begin{figure*}
\centering
\includegraphics[width=6.7in]{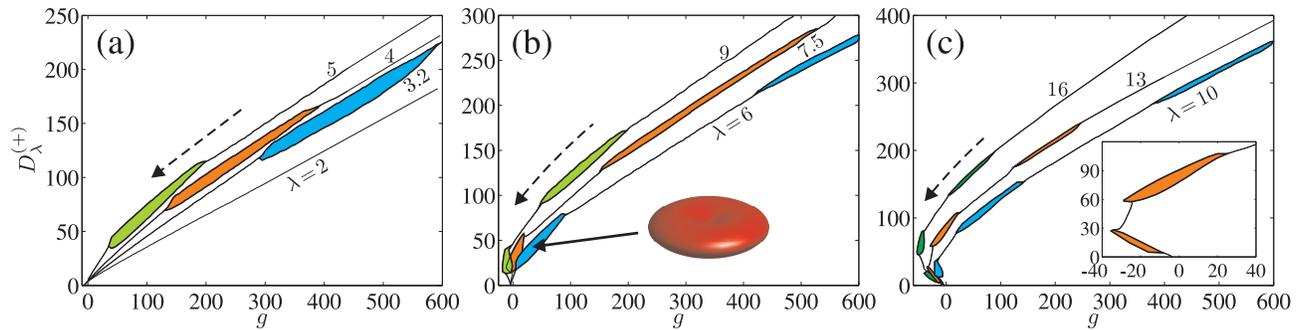}
\caption{The contact interaction strength dependence of the upper critical dipolar interaction $D^{(+)}_\lambda$ for a condensate in pancake-shaped traps. The shaded areas mark the regions of parameter space where the biconcave condensates are found. The broken arrows denote the direction to which the DOIs move as $\lambda$ is increased. The inset of (b) shows the typical isodensity surface of the condensate in DOI. The inset of (c) reveal the detailed structure of the first two DOIs for $\lambda=13$.}
\label{crioblate}
\end{figure*}

This truncation of the dipolar interaction has no physical consequence as long as $R$ is larger than the extend of the actual condensate, but it eliminates the interaction between artificial copies of condensates. This procedure works well in nearly spherical traps. However, for cigar- or pancake-shaped trapping potential, one cannot find a sphere which only contains a single condensate. Here, a natural generalization is to introduce cylindrical truncation on the dipolar interaction potential
\begin{eqnarray}
V_d^{\rm (c)}({\mathbf r})=\left\{\begin{array}{cl}
fd^2(1-3\cos^2\theta)/r^3,&|z|<Z \mbox{ and } \rho<R,\\
0,&{\rm otherwise}.
\end{array}
\right.
\end{eqnarray}
The Fourier transform of $V_d^{\rm (c)}({\mathbf r})$ takes the form
\begin{eqnarray}
\frac{\widetilde V_d^{\rm (c)}}{fd^2}&=& \frac{4\pi}{3}\left(3\cos^2\alpha-1\right)+4\pi e^{-Zk_\rho}\nonumber\\
&&\times\left[\sin^2\alpha\cos(Zk_z)- \sin\alpha\cos\alpha\sin(Zk_z)\right]\nonumber\\
&&-4\pi\int_R^\infty \rho d\rho \int_0^Zdz \cos(k_zz)\nonumber\\
&&\times\frac{\rho^2-2z^2}{(\rho^2+z^2)^{5/2}}J_0(k_\rho\rho),\label{cylcut}
\end{eqnarray}
where $k_\rho=\sqrt{k_x^2+k_y^2}$, $\cos^2\alpha=k_z^2/(k_\rho^2+k_z^2)$, $J_0(\cdot)$ is the zeroth order Bessel function. The first two lines at the r.h.s. of Eq.~(\ref{cylcut}) are previously obtained in Ref.~\cite{ronen}, and the remaining integral term can be easily worked out using numerical integration. With a Gaussian wave function, we have confirmed that the precision of the dipolar interaction energy is improved significantly by introducing cylindrical truncation on $V_d$.

Before we conclude this section, let us briefly summarize the numerical procedure employed to find the ground state wave function of a dipolar condensate. For a given set of control parameters $(\lambda,g,D)$, we first determine the proper values of $R$ and $Z$. Then, using Eq.~(\ref{cylcut}), we numerically compute the Fourier transform of the modified dipolar interaction potential Eq.~(\ref{cylcut}). Finally, following the standard procedure for imaginary time evolution, we obtain the ground state wave function.

\section{Dipolar condensates in a harmonic trap}\label{sresu}
In this section, we shall first map out the stability diagram of a trapped dipolar condensate in the parameter space ($\lambda,g,D$). The value of the trap aspect ratio $\lambda$ ranges from $0.1$ to $20$, which covers both highly prolate and oblate trap geometries. For given parameters $\lambda$ and $D$, increasing the repulsive contact interaction will stabilize the condensate, we therefore manually set an upper limit on $g$ such that $g\leq 600$. The lower limit of $g$, on the other hand, is introduced naturally as the system always becomes unstable by tuning the contact interaction to attractive with sufficiently large strength. Similarly, the lower and upper limits of the dipolar interaction strength are also determined by the stability of the condensate. In the second part of this section, we turn to study the free expansion dynamics of an initially trapped dipolar condensate, aiming at finding signatures of the density oscillation in time-of-flight image.

\begin{figure}
\centering
\includegraphics[width=3.2in]{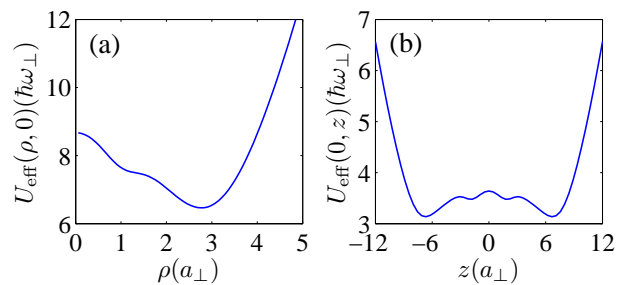}
\caption{The effective potential $U_{\rm eff}(\rho,z)$ of the condensate in the presence of density oscillation for $(\lambda,g,D)=(8,-10,20)$ (a) and $(0.3,200,-49)$ (b).}
\label{effpot}
\end{figure}

\subsection{Stability diagram}
In previous numerical studies~\cite{santos,yi,ronen2}, the stability diagram is plotted in $\lambda$-$D$ parameter plane for a given contact interaction strength $g$ (which is usually taken to be zero). Here, for a fixed trap aspect ratio $\lambda$, we map out the stability diagram in $g$-$D$ parameter plane. As we shall show below, this scheme makes it more convenient for us to understand the behavior of the density oscillation in a trapped dipolar condensate. 

Figure~\ref{cril01} plots the stability diagram in $g$-$D$ parameter plane of a dipolar condensate in a spherical trap, which also represents the basic structure of the stability in other trap geometries. Since the dipolar force is partially attractive, a condensate always becomes unstable as one increases the absolute value of $D$. In fact, for given $g$ and $\lambda$, there exist two critical dipolar interaction strengths, $D_\lambda^{(+)}(g)>0$ and $D_\lambda^{(-)}(g)<0$, such that the condensate is stable only if $D_\lambda^{(-)}(g)<D<D_\lambda^{(+)}(g)$. Alternatively, if $D$ is fixed, the instability can also be induced by decreasing the scattering length (even to negative), similar to what have been done experimentally~\cite{laha,pfau}.

Moreover, when $g$ ($>0$) is so large that the system falls into the Thomas-Fermi regime, the variational calculation shows that the stability boundary for given $\lambda$ only depends on the ratio of $D$ and $g$~\cite{yi}. This properties is clearly reflected in Fig.~\ref{cril01}, as for $g\gtrsim 100$, the critical $D$ values linearly depend on $g$ with slopes $0.255$ and $-0.151$ for $D^{(+)}_\lambda$ and $D^{(-)}_\lambda$, respectively. The reason that the absolute value of $D^{(-)}_\lambda(g)$ is larger than $D^{(+)}_\lambda(g)$ is because, for spherical traps, the dipolar force with negative $D$ is more ``attractive" than that with positive $D$. 

In general, the stability boundaries should depend on the trap aspect ratio $\lambda$. However, we find that, for positive (negative) $D$ in prolate (oblate) trap, the critical dipolar interaction strength is roughly independent of the trap geometry, i.e., $D_{\lambda>1}^{(-)}(g)\approx D_{1}^{(-)}(g)$ and $D_{\lambda<1}^{(+)}(g)\approx D_{1}^{(+)}(g)$. In fact, under these conditions, the condensate shrinks as the dipolar interaction strength grows, such that it only occupies a small spatial volume near the stability boundary. Therefore, the critical dipolar interaction strength becomes insensitive to $\lambda$. For this reason, we shall focus on the upper stability boundary ($D^{(+)}_\lambda$) for oblate traps and the lower stability boundary ($D^{(-)}_\lambda$) for prolate traps.

\begin{figure}
\centering
\includegraphics[width=2.7in]{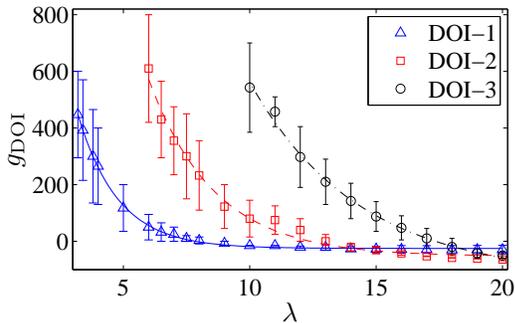}
\caption{The trap aspect ratio dependence of the position and the width (represented by the length of the error bar) of the DOIs. The lines are plotted using Eq.~(\ref{fit}) with fitting parameters listed in Tab.~\ref{tabl}.}
\label{center}
\end{figure}

\begin{table}
\begin{tabular}[b]{c|c|c|c}
\hline\hline
 & $\alpha$ & $\beta$&$\gamma$ \\ 
\hline
DOI-1&$\quad 3559\quad$&$\quad 0.6304\quad$&$\quad 25.19\quad $\\ \hline
DOI-2&6328&0.3856&53.28\\ \hline
DOI-3&6933&0.2312&125.6\\ \hline\hline
\end{tabular}
\caption{The values of fitting parameters for the positions of DOIs.}
\label{tabl}
\end{table}

In Fig.~\ref{crioblate}, we summarize the main results for the stability diagrams in oblate traps. Density oscillations occur when the interaction parameters fall into the shaded areas, namely, the condensate has a biconcave density profile in oblate trap. The origin of the density oscillation can be intuitively understood as follows: the long-ranged dipolar interaction is repulsive in $xy$-plane, therefore, when the dipolar force exceeds certain threshold value the density at the center of the trap is depleted to lower the total energy. On the other hand, as one increases the dipolar interaction strength $D$, the attractive dipolar interaction along $z$-axis also becomes stronger, which destablizes the system. In a slightly oblate trap, the system becomes unstable before such threshold dipolar force required to from the density oscillation is reached. Therefore, for $\lambda\lesssim3.2$, we do not find any density oscillation in condensate wave function for $g$ up to $600$. To get a better understanding of the density oscillation, we plot, in Fig.~\ref{effpot} (a), the effective potential of the system
\begin{eqnarray}
U_{\rm eff}({\mathbf r})=U_{\rm ho}({\mathbf r})+g|\psi({\mathbf
r})|^2+D\int d{\mathbf
r}'V_d({\mathbf r}-{\mathbf
r}')|\psi({\mathbf r}')|^2.\nonumber
\end{eqnarray}
Apparently, when density oscillation occurs in pancake-shaped traps, the effective potential takes a shape of Mexican hat.

\begin{figure}
\centering
\includegraphics[width=2.5in]{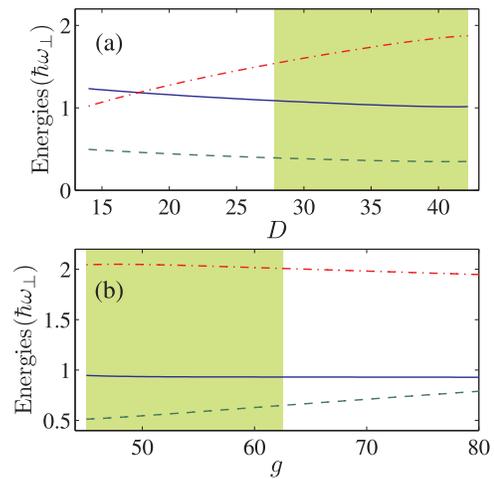}
\caption{(a) Energies as functions of $D$ for $\lambda=6$ and $g=25$. (b) Energies as functions of $g$ for $\lambda=6$ and $D=55$. The solid, dashed, and dash-dotted lines correspond to, respectively, the kinetic, contact interaction, and dipolar interaction  energies. The shaded area marks the range of the interaction parameters where the density oscillation appears.}
\label{energy}
\end{figure}

The first density oscillations island (denoted as DOI-1) appears when $\lambda=3.2$. The shape of the DOI clearly indicates that the stability of the system is slightly increased by forming density oscillation in condensate, and away from DOI, the $D_\lambda^{(+)}$ roughly linearly depends on $g$. As we increase $\lambda$, DOI-1 moves along the stability boundary to negative direction of the $g$-axis. In addition, the area of DOI-1 shrinks. For even larger $\lambda$, the second (DOI-2) and third (DOI-3) density oscillation regions appear. They also move along the stability boundary by following the similar pattern as DOI-1 when the trap aspect ration is increased.

A DOI can be roughly characterized using the contact interaction strength: the range of $g$ parameter covered by the DOI, $\Delta g_{\rm DOI}$, represents its width; while the $g$ value at the center of this range, $g_{\rm DOI}$, is the position of the DOI. Figure~\ref{center} shows the trap aspect ratio dependence of the positions of the three DOIs; the corresponding width is plotted as error bar. One immediately notes that the $\lambda$ dependence of $g_{\rm DOI}$ can be fitted using equation
\begin{eqnarray}
g_{\rm DOI}(\lambda)=\alpha e^{-\beta \lambda}-\gamma,\label{fit}
\end{eqnarray}
where the values of the fitting parameters $\alpha$, $\beta$, and $\gamma$ for different DOIs are listed in Tab.~\ref{tabl}. As shown in Fig.~\ref{center}, the fitted curves agree very well with the numerical results for all DOIs. Moreover, the values of the fitting parameter $\gamma$ also suggest that increasing $\lambda$ will eventually pushes all three DOIs to the region with $g<0$. For large $\lambda$, all three DOIs are squeezed into a small region of $g$ value, but they remain disconnected. We point out that even though Eq.~(\ref{fit}) cannot explain why the DOIs appear under those contact interaction strengths, it can be used to predict the position of DOIs.

Once the position and width of a DOI is known, one may access the DOI by simply increasing the strength of dipolar interaction strength to the value close to the critical value. In Fig.~\ref{energy} (a), we present the typical $D$ dependences of the energies per particle. One immediately notices that the energies vary smoothly as the system enters the DOI, indicating that condensates with and without density oscillation cannot be distinguished from each other by measuring the release energy. As shown in Fig.~\ref{energy} (b), for fixed $D$, one may also enters the DOI by lowering the contact interaction strength. Again, the energies are smooth functions of $g$ across the boundary of a DOI.

\begin{figure}
\centering
\includegraphics[width=2.7in]{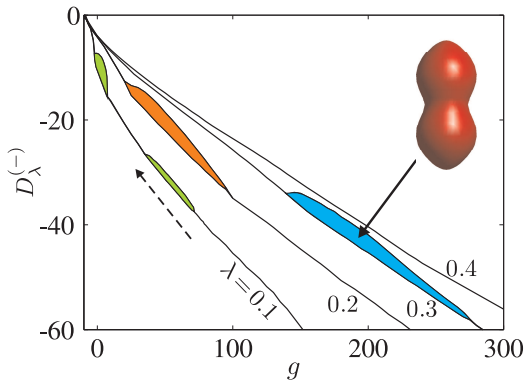}
\caption{The contact interaction strength dependence of the upper critical dipolar interaction $D^{(-)}_\lambda$ for a condensate in cigar-shaped traps. The shaded areas mark the DOIs. The inset shows the typical result for the isodensity surface of the condensate. The broken arrows denote the direction to which the DOIs move as $\lambda$ is decreased.}
\label{cricigar}
\end{figure}

The stability diagram of a dipolar condensate in cigar-shaped traps is presented in Fig.~\ref{cricigar}, where the DOIs appear when $\lambda\lesssim0.3$. In contrast to $D>0$ case, the dipolar interaction becomes repulsive (attractive) along axial (radial) direction for $D<0$. Therefore, along $z$ direction, the repulsive dipolar force depletes the density at the center of the trap such that the isodensity surface of the condensate with density oscillation takes dumbbell-shaped form. Corresponding to dumbbell structure, the effective potential [Fig.~\ref{effpot} (b)] takes the form of a double-well potential. As the trapping potential becomes more pancake-shaped, the DOIs move along the stability boundary to the negative direction of the $g$-axis; in addition, the area of the DOI shrinks.

We remark that our results do not rule out the possibility that there might exist more DOIs for larger $g$ values, since our numerical calculations have only covered limited ranges of the parameters $\lambda$ and $g$.

\subsection{Free expansion}
Now we turn to study the expansion of an initially trapped dipolar condensates. Compared the contact interaction, the anisotropic nature of the dipolar interaction makes the expanded cloud behave quite differently~\cite{yi}. Indeed, it has been used experimentally as a diagnostic tool for the detection the dipolar effects in Bose-Einstein condensate~\cite{stuh,fatt,hulet}.

\begin{figure}
\centering
\includegraphics[width=3.2in]{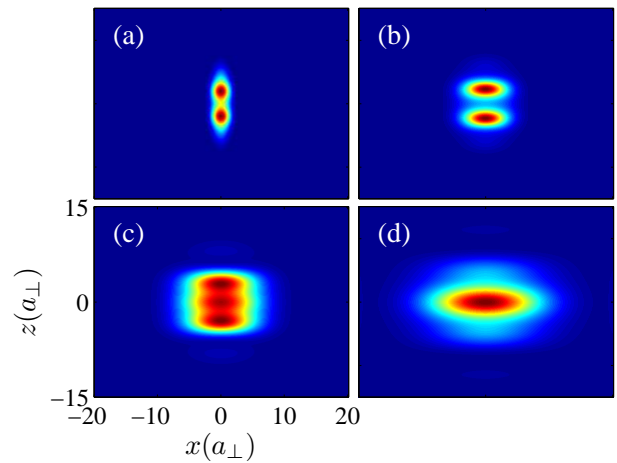}
\caption{The column density of the expanded cloud for $\omega_\perp t=0$ (a), $3$ (b), $5$ (c), and $8$ (d). The initial state are obtained using $(\lambda,g,D)=(0.2,35,-18)$; during the expansion, $g$ is tuned to zero.}
\label{expcigar}
\end{figure}

To study the free expansion dynamics, we consider an initially trapped condensate with control parameters ($\lambda,g,D$) falling into a DOI. At time $t=0$, we switch off the trapping potential, the expansion dynamics of the condensate is then described by Eq. (\ref{gp3d}) with $U_{\rm ho}=0$. If the interactions are ignored completely during expansion, the condensate will expand ballistically such that the time-of-flight image represents the momentum distribution of the trapped condensate. From the Fourier transform of the initial wave function with density oscillation, one can easily deduce that two side peaks would appear in the time-of-flight image, which can be used as the signature of the density oscillation in the initial condensate. 

To demonstrate this, we consider the expansion of an initially cigar-shaped condensate obtained using the control parameter $(\lambda,g,D)=(0.2,35,-18)$. In stead of tuning both $g$ and $D$ to zero, here we only switch off the contact interaction for $t>0$. Figure~\ref{expcigar} shows the column densities of the expanded cloud, i.e.,
\begin{eqnarray}
\bar n(x,z,t)=\int dy n(x,y,z,t).
\end{eqnarray} 
Due to the density oscillation, the initial column density has two peaks, in analog to that of the condensate trapped in a double-well potential. After the condensate expands such that these two peaks overlap, the third peak appears as a result of the interference. At time $\omega_\perp t=5$, three peaks have roughly the same height. When the system evolves continuously, the heights of the two side peaks become lower and lower, but they are still visible at $\omega_\perp t=8$. However, if the contact interaction remains unchanged during free expansion, the above scenario will be spoiled such that the side peaks vanish very quickly for a cigar-shaped condensate.

\begin{figure}
\centering
\includegraphics[width=3.2in]{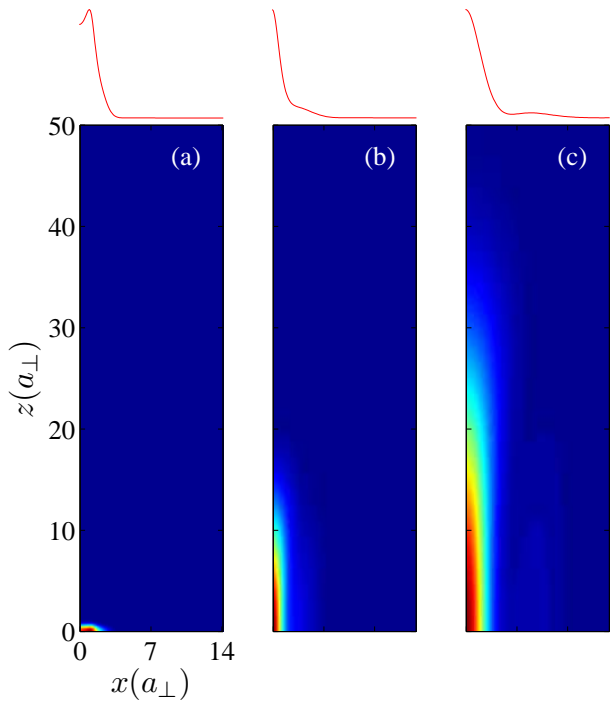}
\caption{The column density of the expanded cloud for $\omega_\perp t=0$ (a), $3$ (b), and $7$ (c). The curves at the upper row show the corresponding column density on $x$-axis, i.e., $\bar n(x,0)$. The initial wave function is obtained using the control parameters $(\lambda,g,D)=(7,-5,33)$.}
\label{expoblate}
\end{figure}

Fortunately, for a pancake-shaped condensate, the side peaks of the column density remain visible for a long time if initially the dipolar interaction is very close to the stability boundary. In Fig.~\ref{expoblate}, we demonstrate the free expansion of a pancake-shaped condensate with initial wave function obtained using the control parameters $(\lambda,g,D)=(7,-5,33)$. To make it easier for visualizing the structure of the expanded cloud, we also plot the column density on $x$-axis. It can be seen that, even for $\omega_\perp t=7$, the side peaks are still distinguishable from the column density, indicating that the time-of-flight image may be used to detect the density oscillation in a dipolar condensate.

\section{Dipolar condensate in a box potential}
\label{box}
In this section we briefly discuss the density oscillations of a dipolar condensate trapped in a box potential. Specifically, we consider a potential which is harmonic along the $z$-axis with frequency $\omega_z$, but is represented by infinite square wells with length $L$ along the $x$- and $y$-axis. Such potential can be realized using, for example, tightly focused light sheet \cite{boxp}. 

\begin{figure}
\centering
\includegraphics[width=3.2in]{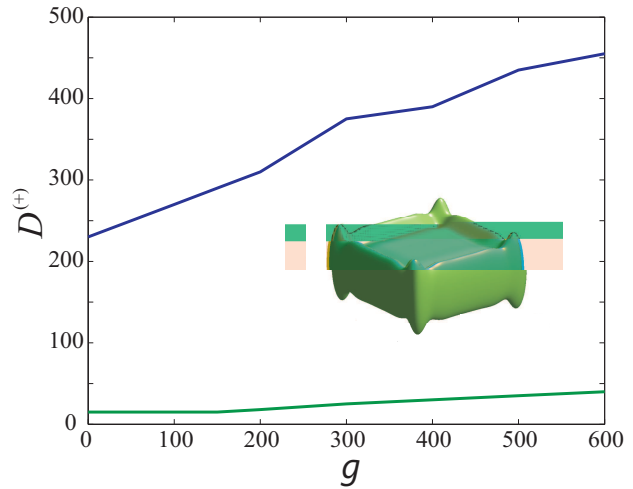}
\caption{The upper solid line represents the upper critical dipolar interaction strength for a dipolar condensate confined in a box potential in the $x$-$y$ plane and a harmonic potential in the $z$ axis. The region between the lower and the upper solid lines represents regions where density oscillation will occur in the ground state. The inset shows an isodensity surface of a dipolar condensate with density oscillation. The length $L$ along the $x$- and $y$-axe is taken to be 28$a_z$, where $a_z=\sqrt{\hbar/(m\omega_z)}$. The dimensionless interaction strengths are defined as $g=4\pi Na/a_z$ and $D=d^2/(\hbar \omega_z a_z^3)$.}
\label{boxf}
\end{figure}

Figure \ref{boxf} displays the stability diagram of such a system. As in the previous case, there exists an critical dipolar interaction strength beyond which the system will collapse. However, a rather significant difference here is the much enlarged region of density oscillation. In fact, here there is no disconnected DOIs and density oscillation dominates a major portion of the parameter space. The inset of Fig.~\ref{boxf} exhibits a typical density plot of a condensate trapped in such a potential. One can easily see the pronounced density oscillations near the walls.

That a hard wall induces density oscillations in dipolar condensate can be understood from the roton excitation of the system \cite{wilson}. Here the effect of the wall is not unlike a vortex line \cite{yi3} or a localized repulsive potential \cite{wilson}: They all create a local superfluid density depression, in the vicinity of which the roton mode becomes soft and manifests itself as density oscillations \cite{wilson}. The same physics also underlies the predicted density ripples in superfluid helium near a boundary \cite{regge} or a vortex line \cite{dalfovo}.

From this study, one can conclude that density oscillations in dipolar condensates can be very easily induced by a rigid wall. However, in case the condensate is not confined in the $x$-$y$ plane (which can be simulated using the periodic boundary condition in the plane), we do not observe any density oscillations --- as dipolar interaction strength increases, the system remains uniform in the transverse plane until it collapses. 

\section{Conclusion}\label{sconc}
To conclude, we have systematically investigated the density oscillations in trapped dipolar condensates by mapping out the stability diagram of the system on $g$-$D$ parameter plane. In addition to the biconcave density profile in an oblate trap, the dipolar interaction can also generate dumbbell-shaped density profiles in prolate traps. On stability diagram, the density oscillation occurs only when the interaction strengths fall into the DOIs. The relation between the positions of DOIs and the trap geometry is explored by fitting the numerical results. We also study the free expansion of the condensate with density oscillation, it is shown that the density oscillation can be identified in the time-of-flight image. Finally, to improve the numerical precision for the calculation of dipolar interaction energy, we introduce the truncated dipolar interaction potential which is nonzero only within a properly chosen cylinder.

This work was supported by NSFC through grants 10974209 and 10935010 and by the National 973 program (Grant No. 2006CB921205). SY acknowledges the support from the ``Bairen" program of Chinese Academy of Sciences. HP acknowledges the support from NSF.


\begin{thebibliography}{}
\bibitem{yi1} S. Yi and L. You, Phys. Rev. A {\bf61}, 041604(R) (2000).

\bibitem{goral} K. G\'{o}ral, K. Rz\c{a}\.{z}ewski, and T. Pfau, Phys. Rev. A {\bf61}, 051601(R) (2000).

\bibitem{santos} L. Santos, G. V. Shlyapnikov, P. Zoller, and M. Lewenstein,
Phys. Rev. Lett. {\bf85}, 1791 (2000).

\bibitem{stuh} J. Stuhler, A. Griesmaier, T. Koch, M. Fattori, T. Pfau, S. Giovanazzi, P. Pedri, and L. Santos, Phys. Rev. Lett. {\bf95}, 150406 (2005).

\bibitem{kurn} M. Vengalattore, S. R. Leslie, J. Guzman, and D. M. Stamper-Kurn,
Phys. Rev. Lett. {\bf 100}, 170403 (2008).

\bibitem{koch} T. Koch, T. Lahaye, J. Metz, B. Frohlich, A. Griesmaier, and T. Pfau, Nat. Phys. {\bf4}, 218 (2008).

\bibitem{laha} T. Lahaye, J. Metz, B. Frohlich, T. Koch, M. Meister, A. Griesmaier, T. Pfau, H. Saito, Y. Kawaguchi, and M. Ueda, Phys. Rev. Lett. {\bf101}, 080401 (2008).

\bibitem{fatt} M. Fattori, G. Roati, B. Deissler, C. D'Errico, M. Zaccanti, M. Jona-Lasinio, L. Santos, M. Inguscio, and G. Modugno,
Phys. Rev. Lett. {\bf 101}, 190405 (2008).

\bibitem{hulet} S. E. Pollack, D. Dries, M. Junker, Y. P. Chen, T. A. Corcovilos, R. G. Hulet, Phys. Rev. Lett. {\bf102}, 090402 (2009).

\bibitem{ye1} S. Ospelkaus, A. Pe'er, K.-K. Ni, J. J. Zirbel, B. Neyenhuis, S. Kotochigova, P. S. Julienne, J. Ye, and D. S. Jin, Nat. Phys. {\bf4}, 622 (2008).

\bibitem{ye2} K.-K. Ni, S. Ospelkaus, M. H. G. de Miranda, A. Pe'er, B. Neyenhuis, J. J. Zirbel, S. Kotochigova, P. S. Julienne, D. S Jin, and J. Ye, Science {\bf 322}, 231 (2008).

\bibitem{goral2} K. G\'{o}ral, L. Santos, and M. Lewenstein, Phys. Rev. Lett. {\bf 88}, 170406 (2002).

\bibitem{yi2} S. Yi, T. Li, and C. P. Sun, Phys. Rev. Lett. {\bf 98}, 260405 (2007).

\bibitem{yi3} S. Yi and H. Pu, Phys. Rev. A {\bf 73}, 061602(R) (2006).

\bibitem{wilson} R. M. Wilson, S. Ronen, J. L. Boh, and H. Pu, Phys. Rev. Lett. {\bf 100}, 245302 (2008).

\bibitem{ronen2} S. Ronen, D. C. E. Bortolotti, and J. L. Bohn, Phys. Rev. A {\bf74}, 013623 (2006).

\bibitem{wilson2} R. M. Wilson, S. Ronen, and J. L. Bohn, Phys. Rev. A {\bf 80}, 023614 (2009).

\bibitem{ronen} S. Ronen, D. C. E. Bortolotti, and J. L. Bohn, Phys. Rev. Lett. {\bf98}, 030406 (2007).

\bibitem{giov} S. Giovanazzi, A. G\"{o}rlitz, and T. Pfau, Phys. Rev. Lett. {\bf89}, 130401 (2002).

\bibitem{santos2} L. Santos, G. V. Shlyapnikov, M. Lewenstein, Phys. Rev. Lett. {\bf90}, 250403 (2003).

\bibitem{pfau} T. Lahaye, T. Koch, B. Fr\"{o}hlich, M. Fattori, J. Metz, A. Griesmaier, S. Giovanazzi, and T. Pfau, Nature {\bf 448}, 672 (2007).

\bibitem{yi} S. Yi and L. You, Phys. Rev. A {\bf 63}, 053607 (2001).

\bibitem{yi4} S. Yi and L. You, Phys. Rev. A {\bf67}, 045601 (2003). 

\bibitem{giov2} S. Giovanazzi, A. Gorlitz, T. Pfau, J. Opt. B: Quantum Semiclassical Opt. {\bf5}, S208 (2003).

\bibitem{giov3} S. Giovanazzi, P. Pedri, L. Santos, A. Griesmaier, M. Fattori, T. Koch, J. Stuhler, and T. Pfau, Phys. Rev. A {\bf 74}, 013621 (2006).

\bibitem{boxp}T. P. Meyrath,
F. Schreck,
J. L. Hanssen, 
C.-S. Chuu, and
M. G. Raizen, Phys. Rev. A {\bf 71}, 041604(R) (2005).

\bibitem{regge}T. Regge, J. Low Temp. Phys. {\bf 9}, 123 (1972).
\bibitem{dalfovo}F. Dalfovo, Phys. Rev. B {\bf 46}, 5482 (1992).

\end{thebibliography}
\end{document}